**Storage of gold nanoclusters in muscle leads to their biphasic *in vivo* clearance**

*Xiao-Dong Zhang,[†] Zhentao Luo,[†] Jie Chen, Hao Wang, Sha-Sha Song, Xiu Shen, Wei Long, Yuan-Ming Sun, Saijun Fan, Kaiyuan Zheng, David Tai Leong\* and Jianping Xie\**

[*] Dr. X. D. Zhang,[†] J. Chen, Dr. H. Wang, S. S. Song, Prof. X. Shen, Dr. W. Long, Prof. Y. M. Sun, and Prof. S. J. Fan
Tianjin Key Laboratory of Molecular Nuclear Medicine, Institute of Radiation Medicine, Chinese Academy of Medical Sciences and Peking Union Medical College, Tianjin, 300192, China
[*] Z. Luo,[†] K. Zheng, Prof. D. T. Leong, and Prof. J. Xie
Department of Chemical and Biomolecular Engineering,
National University of Singapore, 10 Kent Ridge Crescent, 119260, Singapore
Email: chexiej@nus.edu.sg and cheltwd@nus.edu.sg

[ † ] These authors contributed equally to this work.

**ABSTRACT.** Ultrasmall gold nanoclusters (Au NCs for short) show great potential in biomedical applications. Long-term biodistribution, retention, toxicity, and pharmacokinetics profiles are pre-requisites in their potential clinical applications. Here we systematically investigated the biodistribution, clearance, and toxicity of one widely-used Au NC species – glutathione-protected Au NCs or GSH-Au NCs, over a relatively long period of 90 days in mice. We observed that most of the Au NCs were cleared at 30 days post injection (p.i.) with a major accumulation in liver and kidney. However, it is surprising that an abnormal increase of Au amount in the heart, liver, spleen, lung, and testis was observed at 60 and 90 days p.i., indicating that the injected Au NCs formed a V-shaped time-dependent distribution profile in various organs. Further investigations revealed that Au NCs were steadily accumulating in the muscle in the first 30 days p.i., and the as-stored Au NCs gradually released into blood in 30–90 days p.i., which induced a re-distribution and re-accumulation of Au NCs in all blood-rich organs. Further hematology and biochemistry studies showed that the re-accumulation of Au NCs still caused some liver toxicity at 30 days p.i. The muscle storage and subsequent release may give rise to the potential accumulation and toxicity risk of functional nanomaterials over long periods of time.

**1. Introduction**

Gold nanomaterials have attracted exciting attention for medical applications, such as drug delivery, bioimaging, and cancer photothermal or radiation therapy.[1-7] In order for comprehensive clinical acceptance of these gold nanotechnology, the initial few steps require the determination of their pharmacokinetics and *in vivo* clearance.[8-14] Nanomaterials' *in vivo* residence time and their eventual clearance from the circulatory system depends on a web of interwined parameters that are nanomaterial specific and tissue-organ distinctive.[10, 15-16] Larger sized Au nanoparticles (typically above 5 nm) tend to form aggregates in the biological settings due to the protein corona in physiological environments,[17-18] which may result in their preferential accumulation in liver and spleen possibly through nanoparticle induced endothelial leakiness (NanoEL) effect.[19] As highly vascularized organs receive most of the blood in body, circulating nanomaterials can be easily assimilated into these organs and if toxic, can cause organ damage.[17, 20-26] While surface PEGylation can prolong blood circulation time of Au nanomaterials, the as-modified Au nanomaterials may still be trapped in the reticulo-endothelial system (RES) because PEG is still an artificial synthetic molecule and perceived to be foreign by the immune system.[17, 27]

Instead, we suggested that protecting Au nanomaterials with a naturally-occurring biomolecule, such as the ubiquitous tripeptide glutathione (GSH), might be a more natural and therefore a stealthy strategy to evade the RES surveillance and capture. This may overall further improve the stability and residence time of Au nanomaterials in the physiological environment to maximize their therapeutic benefits.[28] In addition, further decrease in the size of Au nanomaterials, especially to below 2 nm, might allow better biomolecule mimicry and interactions with cells and tissues and therefore enhance the stealth mode when under the RES surveillance. Summatively, we combined these two important concepts into a relatively new class of Au nanomaterials – GSH-protected ultrasmall Au nanoclusters (or GSH-Au NCs) and showed in several earlier studies promising biomedical applications.[29-31] These GSH-Au NCs consist of an ultrasmall Au core with sizes below 2 nm enshroud in a self-assembled GSH layer.[32-33] Their biomedical applications have been recently demonstrated in many fields like bioimaging and cancer radiation therapy, showing excellent *in vivo* performance.[29-30]

To further ascertain the clinical suitability of GSH-Au NCs, here we present a multi-parametric *in vivo* study centered on residence times, clearance quantities and rates and toxicity determination of GSH-Au NCs over short- to middle-term (up to 90 days) time points. One recently reported GSH-Au NC – $Au_{29-43}(SG)_{27-37}$ NC was chosen in this study due to its excellent stability and high luminescence.[32] We observed that most of the Au NCs were cleared before 30 days post injection (p.i.) in mice, with a major accumulation in the liver and kidney. However, it is surprising that an abnormal resurgence of Au amount in major blood-rich organs, such as heart, kidney, spleen, liver, lung, and testis was seen after 60 and 90 days p.i., which suggest that the injected Au NCs showed a V-shaped time-dependent dissemination-reaccumulation profile in mice over an extended period of time *in vivo*. Further investigations indicated that the muscle was the major contributor to this unexpected second resurgence. Our data clearly suggest that the *in vivo* distribution and clearance of GSH-Au NCs actually involved two stages instead of the earlier paradigm of a single stage release.[28, 34] In the first stage (1-30 days), the blood circulation and the subsequent renal clearance of GSH-Au NCs made possible, a continuous decrease of Au NCs in those blood-rich organs; what was previously unknown was that there was also a simultaneous accumulation in the muscle most likely due to the NanoEL effect.[19] In the second stage (30-90 days), accumulated Au NCs in the muscle could no longer reside in the muscle and therefore started a continuous release back into the blood system, which then redistributed Au NCs to major blood-rich organs and led to an unexpected increase of Au amount in these organs after 30-90 days p.i.

## 2. Results and Discussion

Luminescent GSH-Au NCs were synthesized and purified according to our published procedures.[32] The as-prepared Au NCs were $Au_{29-43}(SG)_{27-37}$ as determined by the electrospray ionization (ESI) mass spectroscopy. A representative transmission electron microscopy (TEM) image (**Figure 1**a) showed that the GSH-Au NCs had a size below 2 nm. In addition, the as-prepared GSH-Au NCs showed intense orange emission (inset of Figure 1b, under UV illumination), with a distinct emission wavelength at ~610 nm (Figure 1b). The as-prepared GSH-Au NCs had a good stability in solution. No obvious change in luminescence intensity was observed even after 90 days of storage at 4 °C (data not shown). We next monitored the luminescence properties of GSH-Au NCs in the blood plasma over the incubation time. As shown in Figure S1, no obvious change in luminescence intensity was observed in the first 2 h. We only observed a slight decrease of luminescence after 6 h of incubation, and >70% of luminescence intensity of GSH-Au NCs was conserved even after 24 h of incubation. This data clearly suggest that the as-prepared GSH-Au NCs were stable in the blood plasma. While the slight decrease in the luminescence intensity of Au NCs could be due to the formation of protein corona on the surface of GSH-Au NCs, the luminescence dampening is reduced through our smart design of the poly-peptide coating on the Au NCs.

High stability of GSH-Au NCs in blood is crucial for their further *in vivo* applications. This attractive feature also facilitates our *in vivo* biodistribution and clearance study. We chose C57 male mice as the animal model. The as-prepared GSH-Au NCs (3 mM, 0.2 mL) were injected intraperitoneally into the mice (8 mice per group), and the final dose of Au NCs in mice was ~5.9 mg/kg. The mice were sacrificed at 1, 7, 30, 60, and 90 days p.i., and the Au amount in major organs (n = 3) was determined by inductively coupled plasma mass spectrometry (ICP-MS). The biodistribution data were presented in **Figure 2**. The kidney was the dominant organ for Au NCs accumulation, followed by the liver and spleen. Heart, lung, and testis showed relatively low Au concentrations. The detailed biodistribution evolution with time (1, 30, and 90 days p.i.) were illustrated as follows. After the first day p.i., the Au concentrations in kidney, liver, and spleen were 4702 (4% ID/g), 673 (0.6% ID/g), and 1231 (1 % ID/g)) ng/g-tissue, respectively. After 30 days p.i., this measured Au concentration is 509 (0.43% ID/g)) and 652 (0.55 % ID/g)) ng/g-tissue in kidney and liver, respectively; and then 1526 (1.3 % ID/g)) ng/g-tissue Au can be found in spleen. After 90 days p.i., the Au concentration in kidney is 975 (0.8 % ID/g)) ng/g, and 1956 (1.7 % ID/g)) and 1710 (1.4% ID/g)) ng/g in spleen and liver, respectively. The re-increase of Au concentrations in liver and spleen after 90 days p.i. is totally unexpected, and it forms the major focus of this study as the re-accumulation of Au NCs in some organs may directly correlate to their long-term clearance and toxicity issues.

It is well demonstrated that renal clearance is one primary metabolism route for ultrasmall Au NCs with hydrodynamic diameters below 6 nm.[28, 34] Therefore, it is expected that kidney showed the highest Au concentration. On the other hand, since the hydrodynamic diameters of our Au NCs were <3 nm, and particles in this size regime may efficiently escape the RES absorption, a relatively low Au concentration was observed in the liver and spleen after 1 day p.i. The injected Au NCs in mice were slowly cleared *via* the renal clearance pathway, and the total Au amount in blood-rich organs reached the minimum after 30 days p.i. However, our data at 60 and 90 days p.i. suggest that the remaining Au NCs in the body could not be completely cleared during the period of 90 days. These Au NCs have induced an abnormal re-accumulation in those blood-rich organs. The next question we may ask is what is/are the major contributor/s to this abnormal re-accumulation of Au NCs in the body.

To better understand the re-accumulation process, we investigated the time-dependent biodistribution of Au NCs in all major organs. As shown in **Figure 3**, the determined Au amount in liver was increased from 593 to 2224 ng from 1 to 90 days p.i. While kidney is the main organ for foreign particles clearance from the blood, and it is totally expected to observe a drastic decrease in the Au amount from 1081 ng (1 day) to 269 ng (90 days), there was still an unexpected mid term increase resulting in a "V-shape" profile (Figure 3). These "V-shapes" were also observed in the profiles of the lung, heart, and testis with the increase of clearance time – a continuous decrease of Au contents from 1 to 30 days and a continuous or semi-continuous increase from 30 to 90 days. The revived increase of Au contents in the blood-rich organs after 30 days p.i. was unexpected. This data also suggests a biphasic decrease-increase pharmacokinetics profile of Au NCs in mice, in stark contrast to the well known and reported mono-phasic decay pharmacokinetics of nanomaterials.[10, 35-37] Therefore, we surmise that other organs that were not characterized or missed in previous studies might be storing nanomaterials over a long period. This organ may first act as a repository for Au NCs in the first phase, and subsequently serves as a dispensary for the stored Au NCs to other organs, resulting in the re-accumulations of Au NCs in other organs.

We further analyzed the possible reasons for this decrease-increase trend in biodistribution. From the perspective of mass balance, some fluctuations in biodistribution of Au NCs are tolerable in major organs.[28, 36, 38] However, the Au contents in the liver increased two fold to >1500 ng, while the kidney had only about 850 ng decrease of Au NCs, and this is not sufficient to account for the above deficit. More importantly, the kidney is a dominant organ to clear Au NCs in the body, and thus the Au amount decrease in kidney is not attributed solely to the re-distribution between the above mentioned organs. The second possible reason is related to hepato-enteric circulation. It is well known that the hepato-enteric circulation may lead to the storage of foreign particles in the liver and other intestinal organs, which will be released at a later stage.[39] However, if the liver storage is the major contributor to our observations, the Au amount in the liver may undergo a remarkable decrease at some stage, which was inconsistent with the observations in our system – instead the Au amount in the liver increased. Since muscle mass accounts for about 40-60% of the body weight of mice, any small Au amount variation in muscle may lead to an obvious change in other organs. Thus, we hypothesized that muscle might have been the source of the second phase release.

To prove our hypothesis, we measured the time-dependent Au amount in muscle by ICP-MS. As shown in **Figure 4**a, the Au concentration was 2524 ng/g in muscle after 1 day p.i. This value increased to 5588 ng/g after 30 days p.i. A remarkable drop of this value to 187 ng/g was observed in muscle after 90 days p.i. This data provided a direct experimental evidence for our hypothesis, where muscle is the major contributor to the re-accumulation of Au NCs in other organs after 30 days p.i. According to our analysis, the *in vivo* pharmacokinetics of Au NCs in mice could be illustrated in Figure 4b, which involved two stages. The first stage (1-30 days) was the renal clearance and the muscle storage of Au NCs. In this stage, the injected Au NCs in mice were metabolized by kidney and other organs, leading to a continuous decrease of Au concentration in major organs from 1 to 30 days p.i. Meanwhile, possibly due to strong nanoparticle induced endothelial leakiness or NanoEL effect,[19] the Au concentration gradually increased in the muscle, leading to a continuous storage of Au NCs in muscle. The second stage (30-90 days) was the muscle release and redistribution of Au NCs in the blood-rich organs. In this stage, Au NCs stored in the muscle were slowly released and re-entered into the blood circulation. The circulation in the blood therefore redistributed the Au NCs to those blood-rich organs, inducing an abnormal increase of Au NCs in those organs.

The re-accumulation process is interesting, but it could also be risky since for the first time, we showed that there is a longer than the expected residence time of injected nanomaterials in body due to a reflux of nanomaterials from muscle back into the major organs. We then evaluated the toxicological response of Au NCs in the period of 90 days. The dose of Au NCs was set at 5.9 mg/kg, and the body weight, immune response, hematology, and biochemistry of mice were systematically investigated at 1, 7, 30, 60, and 90 days p.i. As shown in Figure S2, in the period of 90 days, the treatment with Au NCs in mice did not induce obvious body weight loss, and spleen and thymus toxicity indices. In addition, we used some gold-standard clinical hematotoxicity markers for the toxicity analysis of injected Au NCs in mice, including the white blood cell (WBC), red blood cell (RBC), hematocrit (HCT), mean corpuscular volume (MCV), hemoglobin (HGB), platelet (PLT), mean corpuscular hemoglobin (MCH), and mean corpuscular hemoglobin concentration (MCHC). The hematology data were presented in **Figure 5**, where no obvious difference in two common indicators (RBC and WBC) was observed in the Au NCs-treated mice compared to those of untreated mice. However, a slight change in MCV, MCH, MCHC, and PLT was observed after 30 days p.i., but these data were recovered after 90 days p.i. This observation was however consistent with the biodistribution data. Although GSH provided a biocompatible layer for the as-prepared GSH-Au NCs, the re-accumulation of them may still induce potential toxicological responses. Moreover, we also examined the standard biochemistry and pathological changes of organs by using immunohistochemistry. As shown in **Figure 6** and **Figure 7**, no obvious organ damages were observed in liver, spleen, and kidney in the period of 90 days after the injection of Au NCs.

The previous studies mostly correlated the potential toxicity of functional nanomaterials to their clearance in body.[13, 40] However, the actual *in vivo* pharmacokinetics of functional nanomaterials could be even more complicated. The data presented in this study clearly indicated that most of the ultrasmall Au NCs can be efficiently cleared by kidney, but the strong nanoEL effect could facilitate the penetration of Au NCs into the relatively poor vascularized tissues, resulting in an appreciable storage of Au NCs in muscle. The as-stored Au NCs can be slowly released into the blood, and the blood circulation may redistribute them to those blood vessels-rich organs. Therefore, the present work suggests the potential adverse risk due to this previously unobserved second wave of accumulation-release of small-sized metal NCs and even other nanomaterials of known toxicities.[41, 42] This new paradigm should be taken into account in future materials design, especially those meant for medical applications.

## 3. Conclusion

In summary, we investigated the long-term (90 days) *in vivo* toxicity and clearance of one commonly-used GSH-Au NCs. It was found that most of Au NCs can be metabolized by renal clearance. Within a typical short test period (30 days), as expected, Au NCs gradually decreased in the highly vascularized organs, such as heart, liver, spleen, and lung. However, at a longer time period (after 30 days), the Au amount in the organs started to climb again. Further investigations confirmed that the muscle behaves as a Au NC sink for the initial 30 days, but subsequently initiates another wave of redistribution to the other organs. The subsequent toxicity test showed that the re-accumulation of Au NCs induced a slight toxicity response at the present dose level. Present work clearly suggests that muscle can store and release ultrasmall nanomaterials, which may cause reaccumulation of functional nanomaterials and result in potential toxicity.

## 4. Experimental Section

*Materials and synthesis:* The orange-emitting $Au_{29-43}(SG)_{27-37}$ NCs were synthesized using a previously reported method. Briefly, freshly prepared aqueous solutions of $HAuCl_4$ (20 mM, 0.50 mL) and GSH (100 mM, 0.15 mL) were mixed with 4.35 mL of ultrapure water at 25 °C. The reaction mixture was heated to 70 °C under stirring (500 rpm) for 24 h. The resultant solution of Au NCs is light yellow under room light and shows strong orange emission under UV (e.g., 365 nm) irradiation. The Au NCs were purified using ultrafiltration [with molecular weight cut off (MWCO) of 3 kDa]. The raw product and purified Au NC solution could be stored at 4 °C for 6 months with negligible changes in their optical properties.

*Materials characterization:* Transmission electron microscopy (TEM) analysis was conducted with a JEOL JEM-2100F microscope operated at 200 kV. The zeta-potential analysis and HD size of the Au NCs was determined with the NanoZS Zetasizer particle analyzer (Malvern). Data were acquired in the phase analysis light scattering mode at 25 °C, and sample solutions were prepared by diluting Au NCs into 10 mM PBS solution (pH 7.0). The UV-vis absorption spectra were recorded on a UV-1800 spectrophotometer (Shimadzu). The photoluminescence (PL) spectra were measured by a fluorescence spectrophotometer (F4600, Hitachi). Stability of Au NCs was evaluated using fluorescence spectra. The Au NCs (3 mM, 05 mL) were diluted one time in human blood plasma, and fluorescence spectra were measured at the time points of 0.5, 2, 6, 12, and 24 hours using a fluorescence spectrophotometer in a 5 mL glass cuvette.

*Animal injection and sample collection*: Animals were purchased, maintained, and handled with protocols approved by the Institute of Radiation Medicine, Chinese Academy of Medical Sciences (IRM, CAMS). 80 Male C57 mice at 11 weeks of age were obtained from IRM laboratories and were housed by 2 mice per cage in a 12 h/12 h light/dark cycle with food and water *ad libitum*. Mice were randomly divided into ten groups (eight mice in each group): 1 day control, 1 day treated, 7 days control, 7 days treated, 30 days control, 30 days treated, 60 days control, 60 days treated, and 90 days control, 90 days treated mice, respectively. The Au NCs (3 mM, 200 μL) was used for the animal experiment using intraperitoneal injection. The concentration was 5.9 mg/kg in each mouse. At every day time point after injection, animals were weighed and assessed for behavioral changes. After 1, 7, 30, and 90 days treatment, all mice were sacrificed, and blood and organs were collected for biochemistry and pathological studies. Mice were sacrificed using isoflurane anesthetic and angiocatheter exsanguination with PBS. One mouse from each group was fixed with 10% buffered formalin following phosphate-buffered saline exsanguination. During necropsy, liver, kidneys, spleen, heart, lung, testis, brain, bladder, and thymus were collected and weighed. To explicitly examine the grade of changes caused by malities, spleen and thymus indexes ($S_x$) were used:

$$S_x = \frac{\text{Weight of experimental organ }(mg)}{\text{Weight of experimental animal }(g)}$$

*Hematology, biochemistry, and pathology*: Using a standard saphenous vein blood collection technique, blood was drawn for hematology analysis (potassium EDTA collection tube). The analysis of standard hematological and biochemical examination was performed. For blood analysis, 1 mL of blood was collected from mice and separated by centrifugation into cellular and plasma fractions. Mice were sacrificed by isoflurane anesthetic and angio catheter exsanguinations, and major organs from those mice were harvested, fixed in 10% neutral buffered formalin, processed routinely into paraffin, stained with H&E and pathology were examined by a digital microscope.

*Biodistribution*: The organs and original solutions from the Au NCs-treated mice (3 samples per group) were digested by using a microwave system CEM Mars 5 (CEM, Kamp Lintfort, Germany). The Au concentration was measured with an ICP-MS (Agilent 7500 CE, Agilent Technologies, Waldbronn, Germany).

*Statistical analyses*: All data presented in this study are the average ± SD of the experiments repeated three or more times. The paired Student's t-test was used for statistical analysis.

ACKNOWLEDGMENT

This work was supported by the National Natural Science Foundation of China (Grant No.81000668), Natural Science Foundation of Tianjin (Grant No. 13JCQNJC13500), the Subject Development Foundation of Institute of Radiation Medicine, CAMS (Grant No.SF1207, SZ1336), and PUMC Youth Fund and the Fundamental Research Funds for the Central Universities (Grant No. 3332013043). Work at National University of Singapore was supported by the Ministry of Education, Singapore, under grants R-279-000-409-112 and R-279-000-350-112.

# Figures and Figures Captions

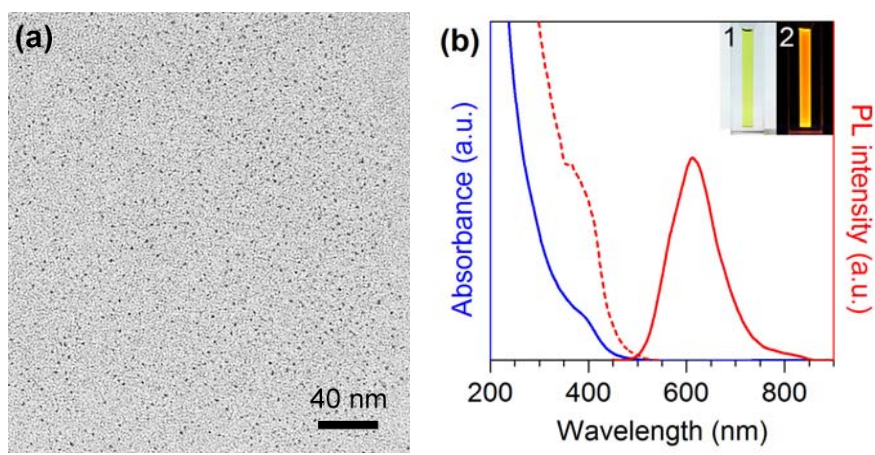

**Figure 1.** (a) TEM image and (b) UV–vis absorption (solid blue line), photoemission (solid red line, $\lambda_{ex}$ = 365 nm), and photoexcitation (dotted red line, $\lambda_{em}$ = 610 nm) spectra of the as-prepared GSH-Au NCs. (Inset) Digital photo of the aqueous solution of the GSH-Au NCs under (1) visible and (2) UV light.

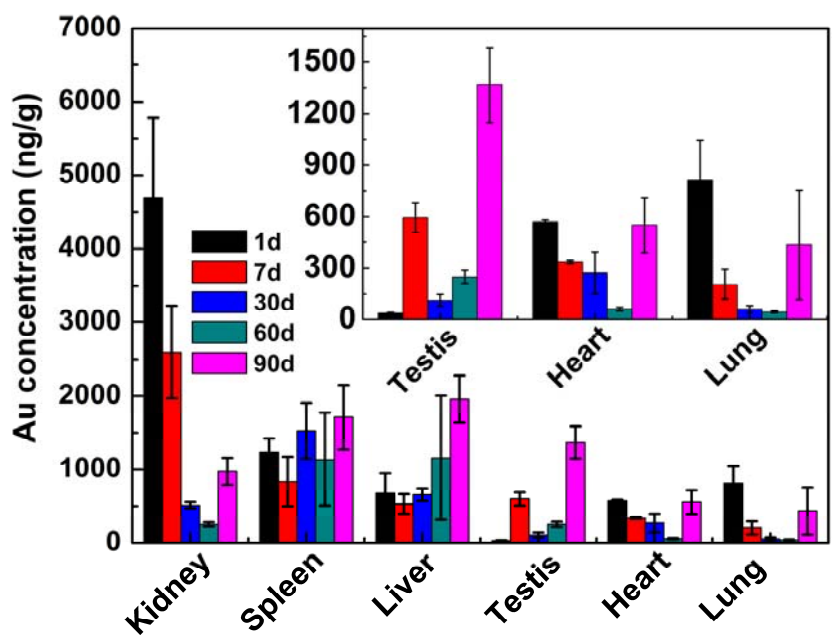

**Figure 2.** Biodistribution of Au NCs determined by ICP-MS at 1, 7, 30, 60, and 90 days p.i.

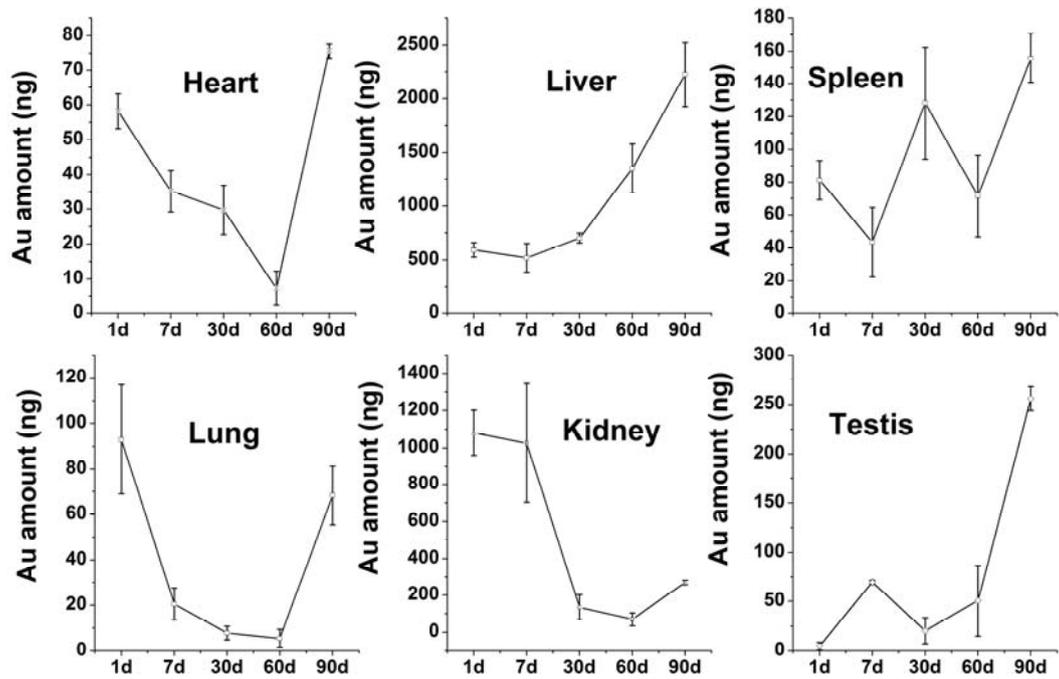

**Figure 3**. Au amount in heart, liver, spleen, lung, kidney, and testis from the mice treated by Au NCs. The data were collected at the different time points of 1, 7, 30, 60, and 90 days p.i.

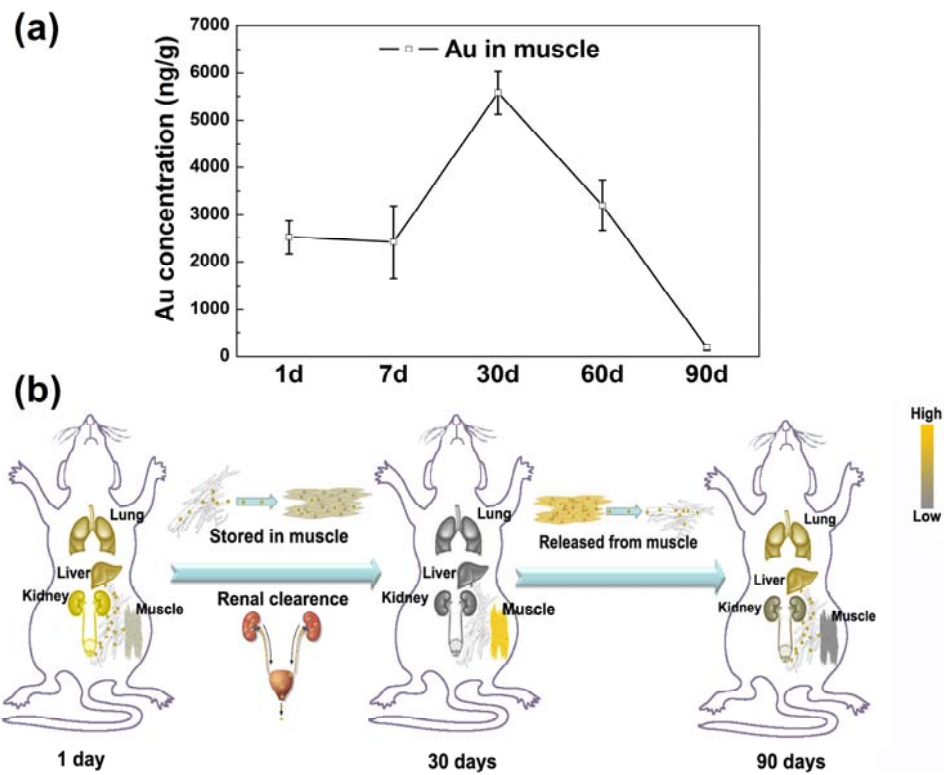

**Figure 4**. (a) Au concentration in muscle from mice treated with Au NCs, and (b) Schematic illustration of time-dependent clearance of GSH-Au NCs injected in mice.

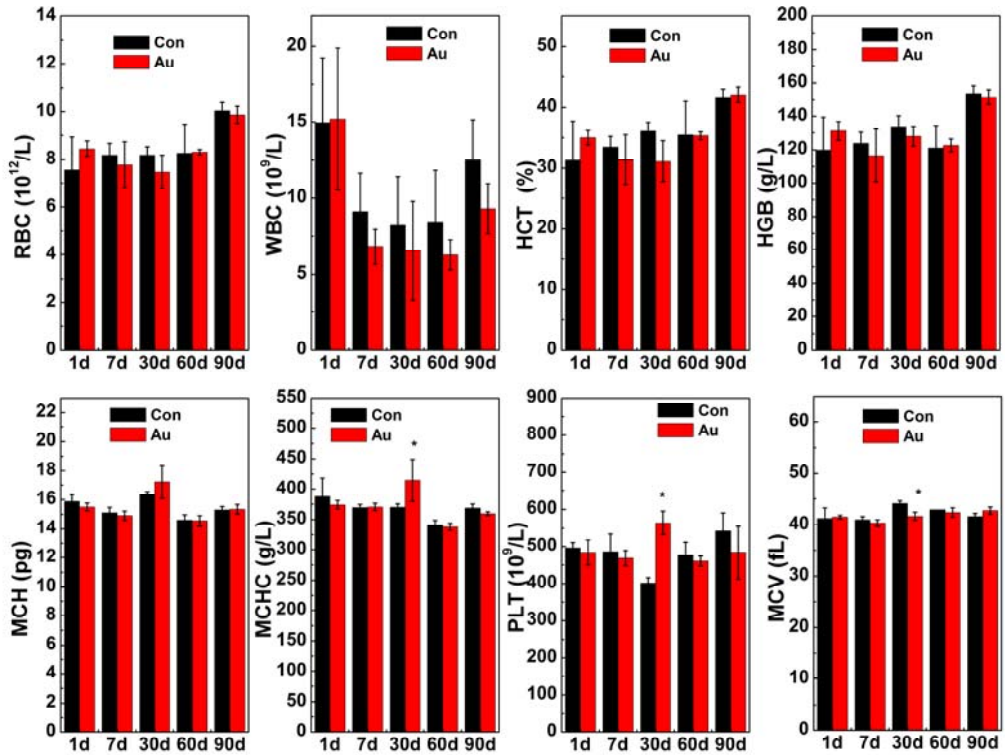

**Figure 5.** Hematology data of the mice treated with the Au NCs. The data were collected at different time points of 1, 7, 30, 60, and 90 days after intraperitoneal injection (5.9 mg/kg). The results show the red blood cells (RBC), white blood cells (WBC), platelets (PLT), mean corpuscular hemoglobin (MCH), mean corpuscular hemoglobin concentration (MCHC), mean corpuscular volume (MCV), hemoglobin (HGB), and hematocrit (HCT). The data were analyzed by Student's t-test and * indicates $p < 0.05$.

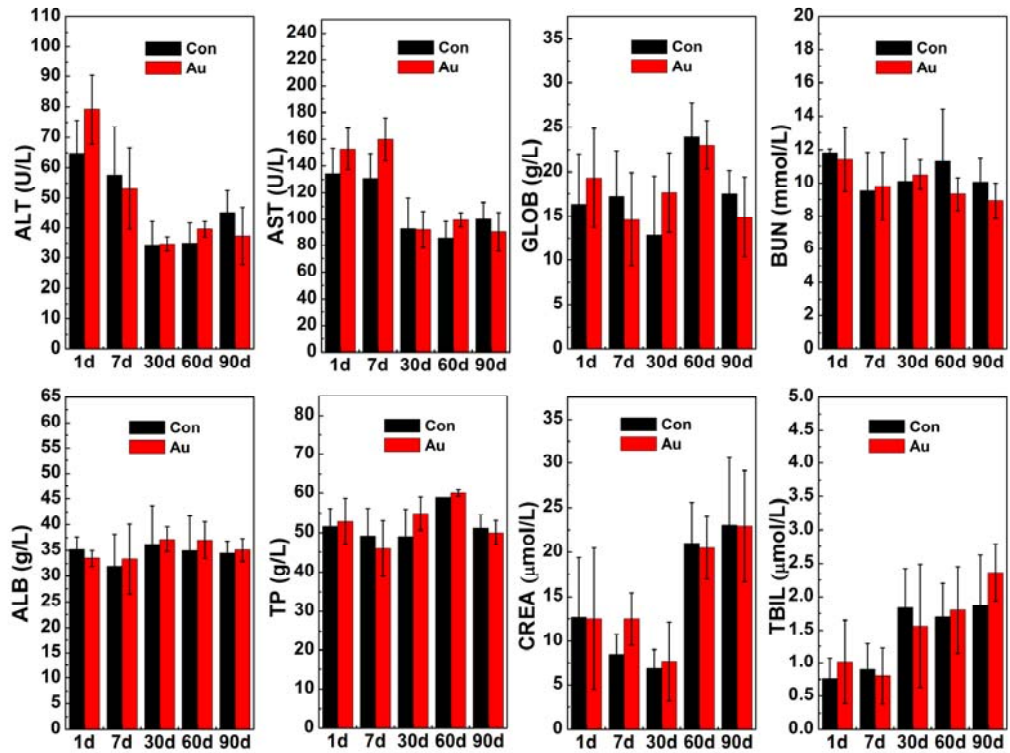

**Figure 6.** Blood biochemistry analysis of the mice treated with the Au NCs at different time points of 1, 7, 30, 60, and 90 days. The results show the mean and standard deviation of aminotransferase (ALT), aminotransferase (AST), total protein (TP), albumin (ALB), blood urea nitrogen (BUN), creatinine (CREA), globulin (GOLB), and total bilirubin (TBIL). The data were analyzed by Student's t-test and * indicates $p < 0.05$.

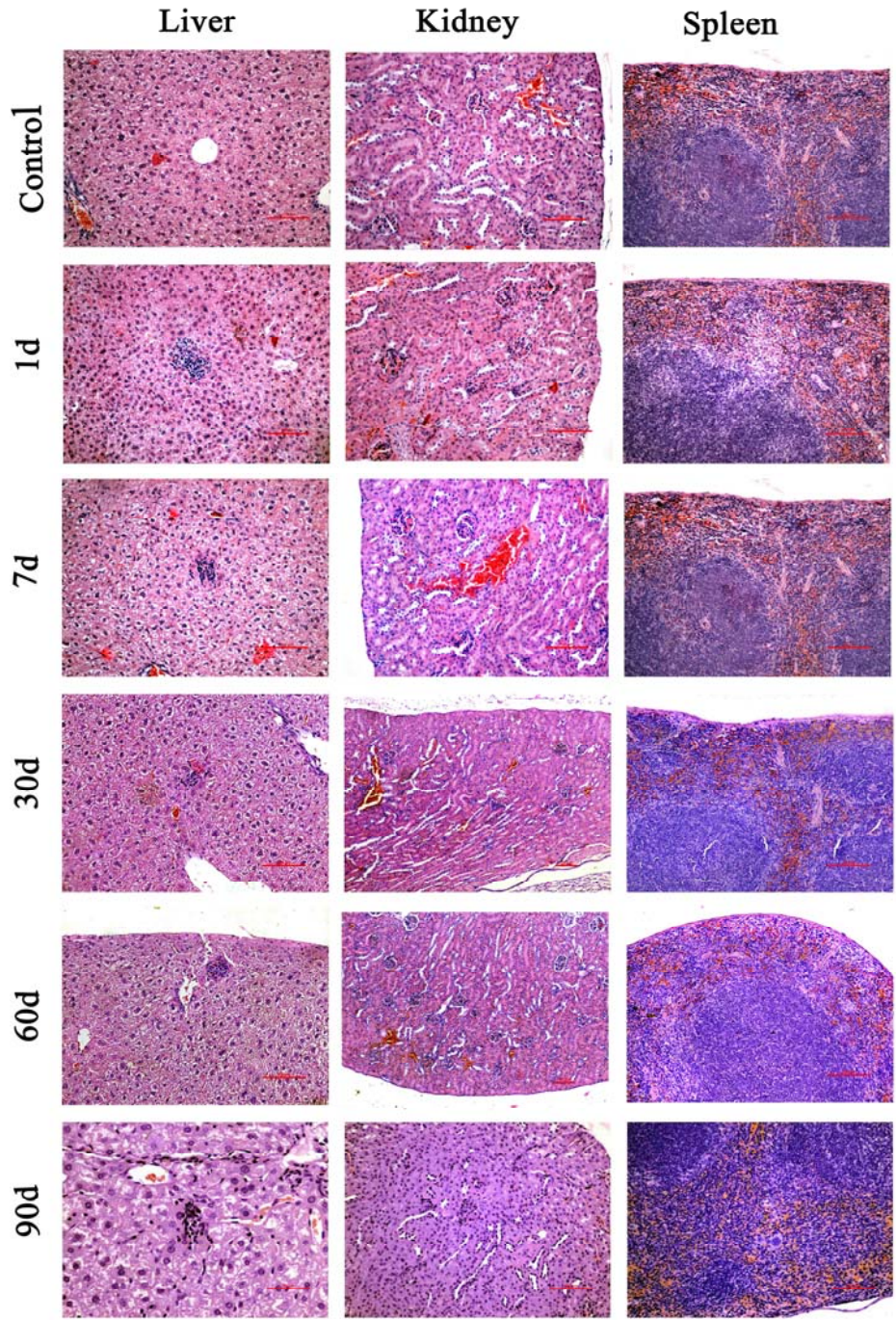

**Figure 7.** Pathological data from the liver, spleen, and kidney of the mice treated with the Au NCs. The data were collected at different time points of 1, 30, and 90 days after injecting 5.9 mg/kg Au NCs.

**GSH-protected ultrasmall Au nanoparticles (1-2 nm) which from a size perspective would normally be cleared expeditiously from the internal organs within a few days post injection, can actually remain and accumulate to very high concentrations for a much longer time (> 60 days) in the muscle versus other internal organs. Subsequently, this accumulation in the muscle starts to release the nanoparticles back into the vascular circulation and initiates another phase of nanoparticle biodistribution.**

**Keyword**

Gold nanoclusters; Biodistribution; Muscle storage; Reaccumulation; Cytotoxicity

**Storage of gold nanoclusters in muscle leads to their biphasic *in vivo* clearance**

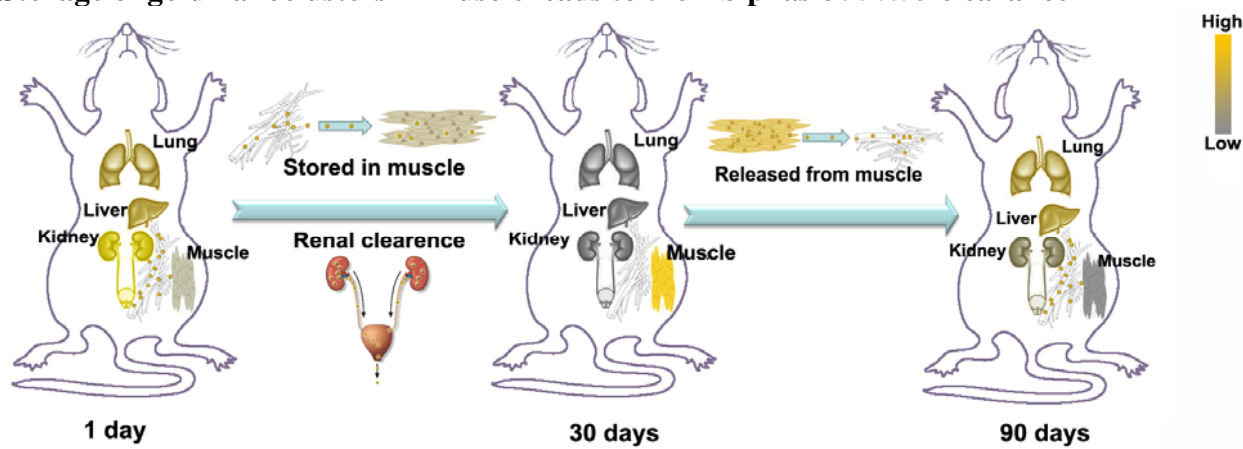

# Supporting Information

**Storage of gold nanoclusters in muscle leads to their biphasic in vivo clearance**

*Xiao-Dong Zhang,[†] Zhentao Luo,[†] Jie Chen, Hao Wang, Sha-Sha Song, Xiu Shen, Wei Long, Yuan-Ming Sun, Saijun Fan, Kaiyuan Zheng, David Tai Leong* and Jianping Xie**

[*]    Dr. X. D. Zhang,[†] J. Chen, Dr. H. Wang, S. S. Song, Prof. X. Shen, Dr. W. Long, Prof. Y. M. Sun, and Prof. S. J. Fan
Tianjin Key Laboratory of Molecular Nuclear Medicine, Institute of Radiation Medicine, Chinese Academy of Medical Sciences and Peking Union Medical College, Tianjin, 300192, China
[*]    Z. Luo,[†] K. Zheng, Prof. D. T. Leong, and Prof. J. Xie
Department of Chemical and Biomolecular Engineering,
National University of Singapore, 10 Kent Ridge Crescent, 119260, Singapore
Email: chexiej@nus.edu.sg and cheltwd@nus.edu.sg

[ † ] These authors contributed equally to this work.

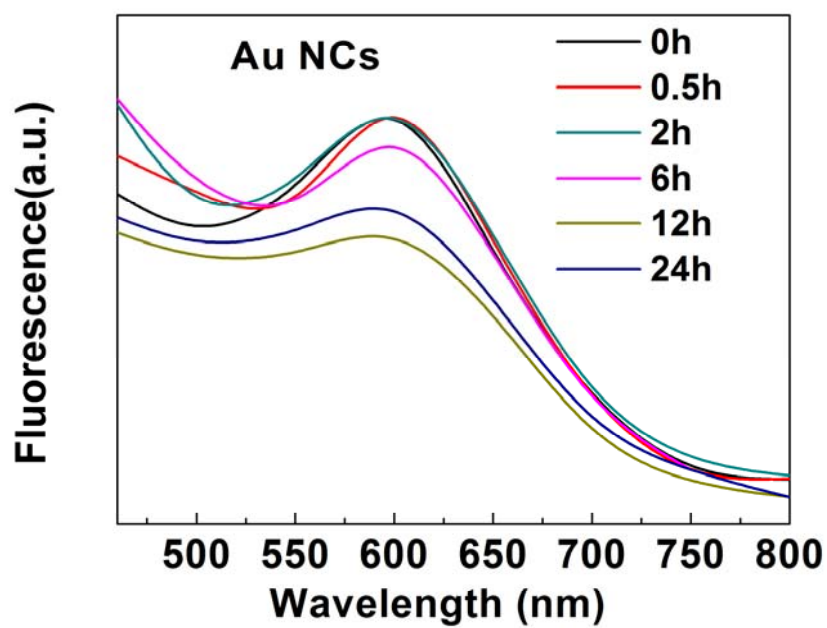

**Figure S1**. Time-dependent fluorescence of the GSH-Au NCs.

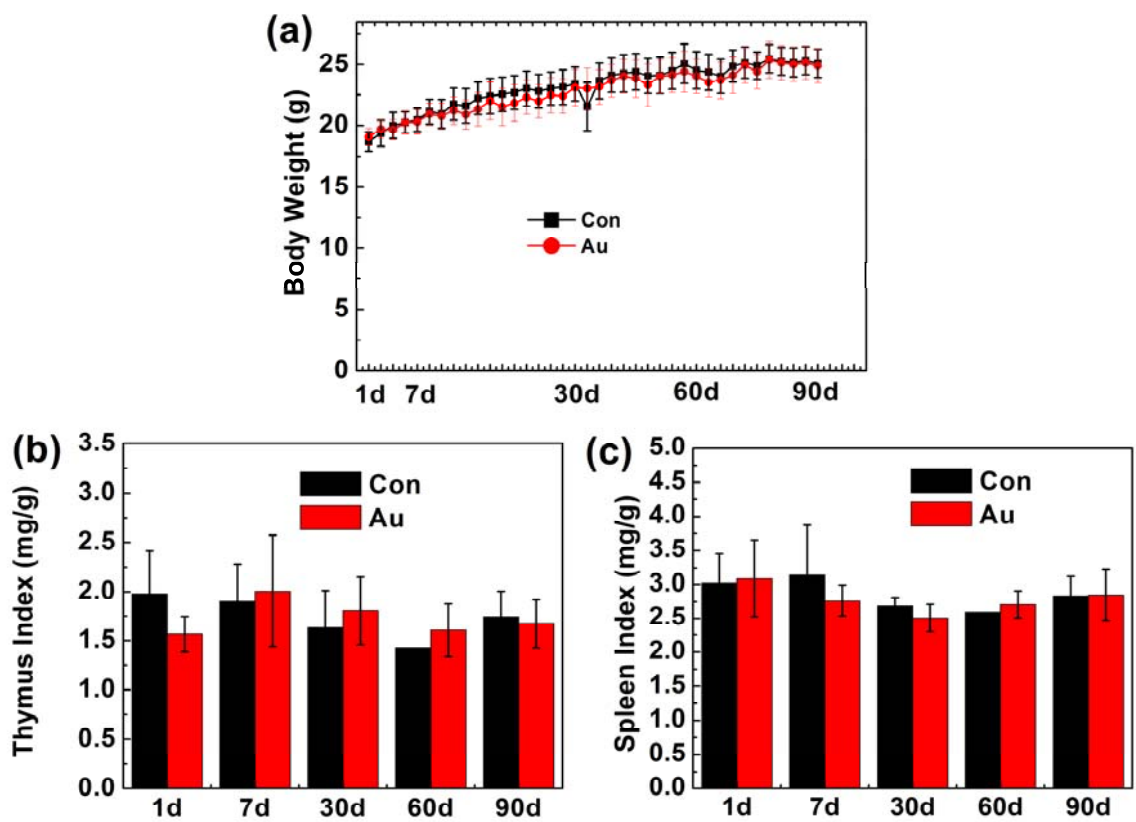

**Figure S2.** (a) Body weight, (b) thymus, and (c) spleen index of the mice treated with the Au NCs.